\documentclass[10pt,letterpaper,twocolumn]{article} 

\usepackage{ol}
\usepackage{amsmath,amssymb}
\usepackage[latin1]{inputenc}
\newcommand{\im}[0]{\text{Im}\,}
\newcommand{\re}[0]{\text{Re}\,}

\begin{document}

\twocolumn[ 

\title{On resolving the refractive index and the wave vector}

\author{Johannes Skaar}

\address{Department of Electronics and Telecommunications, 
Norwegian University of Science and Technology, \\  NO-7491 Trondheim, Norway}


\begin{abstract}
The identification of the refractive index and wave vector for general (possibly active) linear, isotropic, homogeneous, and non-spatially dispersive media is discussed. Correct conditions for negative refraction necessarily include the global properties of the permittivity and permeability functions $\epsilon=\epsilon(\omega)$ and $\mu=\mu(\omega)$. On the other hand, a necessary and sufficient condition for left-handedness can be identified at a single frequency ($\re\epsilon/|\epsilon|+\re\mu/|\mu|<0$). At oblique incidence to semi-infinite, active media it is explained that the wave vector generally loses its usual interpretation for real frequencies.
\end{abstract}

\ocis{000.2690, 260.2110, 350.5500.}
]

After Pendry's perfect lens proposal \cite{pendry2000}, metamaterials and negative refractive index materials have received much attention. The performance of the Veselago--Pendry lens is strongly limited by loss; thus it has been argued that active media may be required to achieve large resolution \cite{ramakrishna2003}.

While the identification of the refractive index of passive media is relatively straightforward, in this Letter I will point out that for general (possibly active) media, one must be careful. Indeed, any direct identification method from $\epsilon$ and $\mu$ at a single frequency, such as the approaches in 
Refs. 3-5,
are incorrect in general. The global behavior of the functions $\epsilon=\epsilon(\omega)$ and $\mu=\mu(\omega)$ must be taken into account to identify the refractive index at a single frequency. 

On the other hand, it is possible to identify a single-frequency condition on $\epsilon(\omega)$ and $\mu(\omega)$ such that the associated medium is left-handed (or right-handed). This means that for active media, there is not necessarily any connection between the sign of the refractive index and the left-handedness/right-handedness \cite{chen2005,chen2006,skaaractivemedia}.

At oblique incidence, another subtle point arises. Let the semi-infinite medium be located in the region $z>0$. Even if the medium is weakly amplifying, it will become clear that in general, the longitudinal wave vector component $k_z$ is not well-defined for real frequencies. In fact, in such media $k_z$ necessarily loses its usual physical interpretation.

To see these properties, we start by defining the refractive index $n(\omega)$ of linear, isotropic, and homogeneous media without spatial dispersion. An implicit time dependence $\exp(-i\omega t)$ is assumed. First, recall that by causality, the medium polarization and magnetization cannot precede the electric or magnetic fields. This means that the functions $\epsilon(\omega)$ and $\mu(\omega)$ are analytic in the upper half of the complex frequency plane \cite{landau_lifshitz_edcm,note1}. Moreover, $\epsilon(\omega)$ and $\mu(\omega)$ tend to +1 as $\re\omega\to\infty$. When $\epsilon(\omega)\mu(\omega)$ has no odd-order zeros in the upper half-plane, $n(\omega)$ for $\im\omega>0$ is defined as the analytic branch of $\sqrt{\epsilon(\omega)\mu(\omega)}$ that tends to $+1$ as $\re\omega\to\infty$. For real $\omega$, $n(\omega)$ is defined as the limit of $n(\omega+i\delta)$ as $\delta\to 0^+$.

When $\epsilon(\omega)\mu(\omega)$ has odd-order zeros in the upper half-plane, $\sqrt{\epsilon(\omega)\mu(\omega)}$ clearly cannot be identified as an analytic function there. In these cases, the refractive index is only defined above the zeros, i.e., for $\im\omega>\gamma$, where $\gamma$ is a positive number \cite{skaaractivemedia}. In fact, then the refractive index cannot be attributed a unique, physical meaning for real frequencies. These media show absolute instabilities in the sense that any small excitation will lead to infinite (or saturated) fields as $t\to\infty$ at a fixed position. Note that this type of instability is different to the convective instabilities in conventional gain media.

One may argue that both signs of the refractive index correspond to valid solutions to Maxwell's equations. However, any physical excitation starts at some time $t=0$. Thus, to see which of the two solutions that is excited, one must use a causal source, such as a unit-step modulated sinusoidal \cite{brillouin}. The monochromatic solution is then found in the limit $t\to\infty$. This calculation may be performed numerically in the time-domain. Alternatively, one can calculate the fields in the frequency domain (or more precisely, Laplace transform domain) as follows \cite{skaaractivemedia}: Consider first a finite slab of thickness $d$, surrounded by vacuum. Then the field at the far end of the slab can trivially be specified, and the field in the slab is determined. By expanding this field into a geometric series, and retaining only the term that yields a nonzero inverse Laplace transform for $t<d/c$, where $c$ is the vacuum light velocity, the resulting field has not felt the presence of the far end. Thus, by subsequently taking the limit $d\to\infty$ the solution in a semi-infinite medium has been obtained. This procedure yields a field of the form $\exp[i\omega n(\omega) z/c]$, where $n(\omega)$ is given by the definition above. 

Note that the definition of $n(\omega)$ is consistent with relativistic causality \cite{brillouin,nussenzveig,skaaractivemedia}. Any other definition of the refractive index must be equivalent to that above to ensure the field dependence $\exp[i\omega n(\omega) z/c]$ in a semi-infinite medium, for a causal plane-wave excitation at $z=0$.

The term ``left-handed medium'', as introduced by Veselago \cite{veselago}, refers to the fact that for simultaneously negative $\epsilon(\omega)$ and $\mu(\omega)$, the electric field, magnetic field, and the wave vector form a left-handed set of vectors. Since $\epsilon(\omega)$ and $\mu(\omega)$ generally are complex, it is common to rather adopt the following definition: A medium is said to be left-handed at the frequency $\omega$ if the associated, time-averaged Poynting vector and the phase velocity point in opposite directions. If they point in the same direction, the medium is right-handed.

Having established the definitions, we now turn to the conditions for negative refraction. For simplicity the $\omega$ dependence in the notations will be omitted. First we treat the well-known case where the medium is passive, i.e., $\im\epsilon$ and $\im\mu$ are positive for positive frequencies. According to the definition of the refractive index, $n\to +1$ as $\re\omega\to\infty$ in the upper half-plane. Since $\epsilon, \mu\to +1$ in this limit, $n\to+1$ is achieved by letting
\begin{equation}
\label{nfromepsilonmupassive}
n=\sqrt{|\epsilon||\mu|}\exp[i(\arg\epsilon+\arg\mu)/2],
\end{equation}
where the complex arguments are restricted to the interval $(-\pi,\pi]$ (i.e., $[0,\pi]$ for positive frequencies). It is well known that for passive materials, $\epsilon$ and $\mu$ cannot take any real value less than or equal to zero, in the upper half-plane \cite{landau_lifshitz_edcm}. (In particular, $\epsilon\mu$ is zero-free there.) Thus, with the above restrictions on the complex arguments, the right-hand side of Eq. \eqref{nfromepsilonmupassive} becomes analytic in the upper half-plane and represents the correct $n$.

With the expression \eqref{nfromepsilonmupassive} it is now straightforward to obtain the conditions for negative refraction: The real part of the refractive index of a passive medium is negative if and only if \begin{equation}
\label{condnegrefrpassive}
\arg\epsilon + \arg\mu>\pi.
\end{equation}
Inequality \eqref{condnegrefrpassive} is equivalent to the conditions that appeared in 
Refs. 14 and 15.

The condition for left-handedness is found by calculating the time-averaged Poynting vector. Assuming a plane wave of the form $\exp(i\omega nz/c)$, we easily find that the sign of the Poynting vector is given by that of $\re(n/\mu)$. With the help of Eq. \eqref{nfromepsilonmupassive} (which implicitly assumes signal front velocity in the $+z$-direction), we find that $\re(n/\mu)=\sqrt{|\epsilon|/|\mu|}\cos[(\arg\epsilon-\arg\mu)/2]$ cannot be negative. Physically this means that a causal excitation of a wave with signal front velocity in the $+z$-direction leads for a passive material necessarily to energy flow in the same direction. The result can also be rephrased as follows: A passive material is left-handed if and only if $\re n<0$, i.e., if and only if condition \eqref{condnegrefrpassive} is fulfilled.

For general (possibly active) media, the situation is completely different. We limit ourselves to media with no absolute instabilities, i.e., media for which $\epsilon\mu$ has no odd-order zeros in the upper half-plane. Then the refractive index can be identified as an analytic function in the upper half-plane. An analytic function is of course continuous; thus the refractive index is given by 
\begin{equation}
\label{nfromepsilonmu}
n=\sqrt{|\epsilon||\mu|}\exp[i(\varphi_\epsilon+\varphi_\mu)/2],
\end{equation}
where $\varphi_{\epsilon}+\varphi_{\mu}$ is the complex argument of $\epsilon\mu$, unwrapped such that $2\pi$ discontinuities are removed, and such that it tends to zero as $\re\omega\to\infty$. Strictly, for real frequencies, Eq. \eqref{nfromepsilonmu} is evaluated in the limit $\im\omega\to 0^+$. However, usually $\epsilon\mu$ is continuous and zero-free at real frequencies except possibly at $\omega=0$. Then Eq. \eqref{nfromepsilonmu} can be used directly for real frequencies, by unwrapping $\varphi_\epsilon+\varphi_\mu$ for $\omega>0$ and $\omega<0$, and ensuring the limit $\varphi_\epsilon+\varphi_\mu\to 0$ for $\omega\to\pm\infty$. Note that in these common cases $n(\omega)$ is continuous for real $\omega>0$. Also note that the phase unwrapping procedure means that the sign of the refractive index at a certain frequency is dependent on the global properties of the function $\epsilon(\omega)\mu(\omega)$. 

While it is impossible to establish a criterion for negative refraction that only considers a single frequency, one can identify a general condition for left-handedness. Let $\tilde n=\sqrt{|\epsilon||\mu|}\exp[i(\arg\epsilon+\arg\mu)/2]$, where the complex arguments are restricted to the range $(-\pi,\pi]$. ($\tilde n$ is not necessarily equal to the physical refractive index $n$ as determined by Eq. \eqref{nfromepsilonmu}. However, since $\tilde n=\pm n$ we can still determine left-handedness/right-handedness by comparing the signs of $\re\tilde n$ and $\re(\tilde n/\mu)$.) If $\re\tilde n>0$, we find that $\re(\tilde n/\mu)>0$ when $|\arg\epsilon-\arg\mu|<\pi$. If $\re\tilde n<0$, $\re(\tilde n/\mu)>0$ always. This means that the medium is right-handed when $|\arg\epsilon+\arg\mu|<\pi$ and $|\arg\epsilon-\arg\mu|<\pi$, and left-handed otherwise. Noting that $\re\epsilon/|\epsilon|+\re\mu/|\mu|=\cos\arg\epsilon+\cos\arg\mu=2\cos\frac{\arg\epsilon+\arg\mu}{2}\cos\frac{\arg\epsilon-\arg\mu}{2}$, this can be rephrased as follows: A medium is left-handed if and only if
\begin{equation}
\label{LHcond}
\re\epsilon/|\epsilon|+\re\mu/|\mu|<0.
\end{equation}
The condition \eqref{LHcond} appeared previously in 
Ref. 15;
however note that the alternative condition in 
Ref. 15,
$\re\epsilon\,\im\mu + \re\mu\,\im\epsilon<0$, and the condition in 
Ref. 14 
only apply to passive media.

The subtle behavior of certain active media can be illustrated by the following example: Let $\epsilon_\text{p}(\omega)$, $\mu_\text{p}(\omega)$, and $n_\text{p}(\omega)$ be the electromagnetic parameters of a passive material. At a certain frequency $\omega_1$ we assume that the medium is left-handed, with $\epsilon_\text{p}=\mu_\text{p}=-1+i\alpha$, where $0<\alpha\ll 1$. Thus the refractive index at $\omega_1$ is $n_\text{p}=-1+i\alpha$. Now, consider an active material with permittivity $\epsilon(\omega)=\epsilon_\text{p}(\omega)\mu_\text{p}(\omega)$ and permeability $\mu(\omega)=1$. Since $\epsilon(\omega)$ is analytic in the upper half-plane, and has the correct asymptotic behavior, this active medium is causal and realizable, at least in principle. From the definition, we find immediately the refractive index $n(\omega)=n_\text{p}(\omega)$. Thus, at $\omega_1$ the refractive index is $n=-1+i\alpha$ while $\epsilon\approx 1-2i\alpha$ and $\mu=1$. In other words, at $\omega_1$, although the medium has permittivity and permeability identical to that of conventional, positively refracting gain media, the refractive index is negative. (The fact that a conventional gain medium with e.g. inverted Lorentzian susceptibility yields positive $\re n$ is easily obtained using the definition of the refractive index.) The active, negative index medium above is clearly right-handed, and at $\omega_1$ both phase velocity and Poynting's vector point towards the source. The properties of similar media are discussed in 
Refs. 6 and 8.

At oblique incidence, the interesting question is the sign of $k_z=\sqrt{n^2\omega^2/c^2-k_x^2}$ rather than the sign of $n$. Here the transversal wave number $k_x$ is assumed to be real. First we assume that $n^2\omega^2/c^2-k_x^2\neq 0$ everywhere in the upper half-plane. This is always the case for passive media \cite{note2}; however, as will be discussed below there are active media for which the condition does not hold. By causality, $k_z$ is identified as the analytic function of $\omega$ in the upper half-plane that tends to $+\omega/c$ as $\re\omega\to\infty$. It follows \cite{note3} that $k_z$ must be a continuous function of $k_x$. Since the sign of $\im k_z^2$ is independent of $k_x$, we can conclude that variation of $k_x$ does not alter the quadrant of $k_z$. In other words, the signs of $\re k_z$ and $\im k_z$ for any real $k_x$ are equal to those of $\re n$ and $\im n$, respectively. Physically this means that if $\re n<0$, we will get negative refraction at a boundary to vacuum, independent of the angle of incidence. Also, if a wave is damped at normal incidence, it is also damped at any other angle. For large $k_x$ corresponding to evanescent waves in vacuum, we find that the waves remain evanescent decaying in semi-infinite, passive materials. This was also noted by Pendry \cite{pendry2000}.

It is tempting to conclude that if a wave is amplified at normal incidence, it will also be amplified for large $k_x$, corresponding to evanescent waves in vacuum. However, in the previous paragraph we assumed that $n^2\omega^2/c^2\neq k_x^2$ in the upper half-plane. In fact, most conventional gain media do not satisfy this requirement. Indeed, consider a weakly amplifying, inverted Lorentzian medium with $\epsilon(\omega)=1-f(\omega)$ and $\mu(\omega)=1$, where 
\begin{equation}
\label{lorentz}
f(\omega)=\frac{F\omega_{0}^2}{\omega_{0}^2-\omega^2-i\omega\Gamma}.
\end{equation}
Here, $\omega_{0}$, $\Gamma$, and $F$ are positive parameters. Assuming $k_x\neq 0$ and small gain ($F\ll \Gamma |k_x|c/\omega_0^2$ or $F\ll |1-k_x^2c^2/\omega_0^2|$), we find that $n^2\omega^2/c^2=k_x^2$ is satisfied for $\omega=|k_x|c(1+a)$, for a certain $a$ with $\im a>0$ and $|a|\ll 1$. Thus, although $n$ has no branch points in the upper half-plane, $k_z$ may have branch points there. In such cases $k_z=k_z(\omega)$ clearly cannot be identified as an analytic function in the upper half-plane. The necessary branch cuts mean that $k_z$ loses its usual interpretation for real frequencies. The physical, time-domain electric field is found by a Bromwich integral above the branch cuts. Alternatively, one may integrate along the real frequency axis (inverse Fourier transform); however then contour integrals around the branch cuts in the upper half-plane must be added to the result. These integrals blow up with time and imply an instability as seen from a fixed position $z$. This result can be interpreted as follows: For nonzero $k_x$, any causal excitation involves necessarily the frequency where $k_z=0$. This wave propagates an infinite distance along the transversal direction before arriving at the position $z$; thus it picks up an infinite amount of gain (in the absence of saturation).

Finally, we note that for active media, no fundamental principle prevents $n$ from being purely imaginary with $\im n<0$, even in a semi-infinite, causal medium. For example, consider the causal medium $\epsilon(\omega)=[1-f(\omega)]^2$ and $\mu(\omega)=1$, where $f(\omega)$ is given by Eq. \eqref{lorentz}. Using the definition of the refractive index, we obtain $n(\omega)=1-f(\omega)$. Assuming $F=1$ and $\Gamma<\omega_0$, we find $n=-i\sqrt{\omega_0^2-\Gamma^2}/\Gamma$ at the frequency given by $\omega=\sqrt{\omega_0^2-\Gamma^2}$. Thus, using e.g. a unit-step modulated sinusoidal source at $z=0$ (plane wave at orthogonal incidence), the monochromatic field in the limit $t\to\infty$ will be an increasing exponential as a function of $z$. This result is in contrast to the statement in 
Refs. 3 and 4,
where $n=-i\alpha$, $\alpha>0$ is claimed to be physically unaccessible in semi-infinite media. The medium will certainly saturate at large distances away from the source, but this is not fundamentally different from the situation with amplified, propagating waves.

In conclusion, although the condition \eqref{LHcond} for left-handedness is valid even for active media, it is impossible to give conditions for negative refraction that consider $\epsilon$ and $\mu$ at a single frequency. In general it is impossible to choose the right sign of $n$ based on the single-frequency Poynting vector. For nonzero transversal wave number, one must be particularly careful: In many active media (e.g. practical low-gain media), the longitudinal wave vector $k_z$ is not well-defined for real frequencies. In passive media and the few active media where $k_z$ is meaningful for real frequencies, the signs of $\re k_z$ and $\im k_z$ are identical to those of $\re n$ and $\im n$, respectively.
 

\begin{thebibliography}{10}
\bibitem{note1}
Writing $\epsilon(\omega)=1+\int_0^\infty x(t)\exp(i\omega t)\text{d}t$ (and similarly for $\mu(\omega)$), we restrict ourselves to stable media in the sense that the response function $x(t)$ is bounded.
\bibitem{note2}
For passive media $\im[n(\omega)\omega-\omega]>0$ for real frequencies. As the function $\im[n(\omega)\omega-\omega]$ is harmonic, Poisson's integral formula ensures that $\im[n(\omega)\omega-\omega]>0$ remains valid in the upper half-plane. It follows that $n(\omega)\omega$ cannot be real there.
\bibitem{note3}
If the function $k_z(\omega,k_x)=\sqrt{n^2(\omega)\omega^2/c^2-k_x^2}$ were discontinuous in $k_x$, we could find a $(\omega,k_x)$ and a tiny $\delta>0$ such that $k_z(\omega,k_x+\delta)\approx-k_z(\omega,k_x)$. This leads to a contradiction since $k_z(\omega,k_x)$ is continuous in $\omega$, and $k_z(\omega,k_x+\delta)\to k_z(\omega,k_x)$ as $\re\omega\to\infty$.
\end{thebibliography}

\end{document}